\documentclass[prd,aps,twocolumn,showpacs]{revtex4} 
\usepackage{amssymb}
\usepackage{graphicx}

\begin{document}

\draft
\title{ Dynamical Foundations of the Brane Induced Gravity}  
\author{			Keiichi Akama}
\address{	Department of Physics, Saitama Medical University,
 			 Saitama, 350-0495, Japan}
\author{			Takashi Hattori}
\address{	Department of Physics, Kanagawa Dental College,
 			 Yokosuka, 238-8580, Japan}
\date{\today}

\begin{abstract}
We present a comprehensive formalism to derive precise expressions 
	for the induced gravity of the braneworld,
	assuming the dynamics of the Dirac-Nambu-Goto type.
The quantum fluctuations of the brane at short distances
	give rise to divergences,
	which should be cutoff at the scale of the inverse thickness of the brane.
It turns out that the induced-metric formula 
	is converted into an Einstein-like equation 
	via the quantum effects.
We determine the coefficients of the induced cosmological and gravity terms,
	as well as those of the terms including 
	the extrinsic curvature and the normal connection gauge field.
The latter is the characteristic of the brane induced gravity theory,
	distinguished from ordinary none-brane induced gravity.
\end{abstract}

\pacs{ 04.50.-h, 04.62.+v, 11.25.-w, 12.60.Rc}




\maketitle

\section{Introduction}

General relativity is based on the premises 
that the spacetime is curved affected by matter 
	according to the Einstein equation, 
	and 
that the objects move along the spacetime geodesics.
The gravitations are apparent phenomena 
	of the inertial motions in the curved spacetime. 
This successfully explains 
	why the motion in the ``gravitational" field is universal, 
	i.e. blind to the object properties (mass, charge etc.), 
	and why the ``gravitational" forces 
	are subject to the Newton's law of gravitation
	(within the ordinary precision). 
It is supported by further precise observations.
It raises, however, another fundamental question 
	{why the spacetime is curved so as the Einstein equation indicates.}
A possible answer was as follows: 
Suppose our spacetime is a 3+1 dimensional embedded object 
	like domain walls or vortices
	(braneworld, \cite{Fronsdal}--\cite{Jardim:2011gg}) 
	in higher dimensional spacetime,
	and the gravity is induced through quantum fluctuations
	(brane induced gravity, 
	\cite{Akama78}, \cite{Akama82}, \cite{Pavsic},
	\cite{Akama87}--\cite{Akama88b},
	\cite{BIG}, \cite{DvaliGabadadze}, 
	\cite{AH}, \cite{Krause:2008sj}).
The cosmological constant should be finely tuned.
Then, the brane-gravity field emerges as a composite,
	which obeys an Einstein-like equation 
	at least at low curvatures,
	just like in the case of the (non-brane) induced gravity theory
	\cite{Sakharov}--\cite{indg3}.

The ideas of the braneworld and the brane induced gravity 
	have been studied extensively in the last three decades. 
Physical models were constructed with topological defects 
	in higher dimensional spacetime \cite{Akama82}--\cite{Akama88b},
	and they were realized as ``D-branes" in the superstring theory 
	\cite{DaiLeighPolchinski}--\cite{HoravaWitten}.
They were applied to the hierarchy problem 
	with large extra dimensions, or with warped extra dimensions 
	\cite{Antoniadis}, \cite{ADD}--\cite{RS}.
It was argued that the brane induced gravity would imply
	infrared modifications of the gravity theory
	\cite{GregoryRubakovSibiryakov}--\cite{DvaliGabadadze}.
The ideas have been studied in wide areas including
	basic formalism, \cite{Th1}--\cite{Th4},
	brane induced gravity \cite{AH}--\cite{Krause:2008sj},
	particle physics phenenomenology \cite{Ida:2002ez}--\cite{SarrazinPetit},
	and cosmology \cite{KantiKoganOlivePospelov}--\cite{Jardim:2011gg}
	with many interesting consequences.

In this paper, 
	we establish a precise formalism to derive the expressions 
	for the quantum induced effects on the brane.
For definiteness, we follow 
	the simplest model with the Dirac-Nambu-Goto type dynamics 
	\cite{DiracNambuGoto}.
Such a model was first considered in \cite{Akama78} with scalar fields which
	we now interpret as the position coordinate of the brane in higher dimensions.
Such an interpretation motivated us to consider the model of braneworld
	as a topological object moving in higher dimensions in \cite{Akama82}.
The idea of the brane induced gravity has been considered 
	repeatedly in the literature.
They were, however, more or less naive both in model setting and derivation.
Here we present a comprehensive formalism from its foundation to the precise outcomes.
Among the quantum fluctuations,
	the only meaningful ones are those transverse to the brane.
The quantum loop effects are divergent, 
	which should be cutoff 
	at the scale of the inverse thickness of the brane.
We adopt the regularization scheme used in the original work \cite{Akama78},
	and calculate them at the one-loop level.
We determine the coefficients of the induced cosmological and gravity terms,
	as well as those of the terms including 
	the extrinsic curvature and the normal connection gauge field
	\cite{GibbonsSuggest}.
The induction of the latter terms 
	is characteristic of the brane induced gravity theory,
	distinguished from the ordinary (non-brane) induced gravity.
It turns out that the induced-metric formula 
	is converted into an Einstein-like equation 
	via the quantum short-distance effects.

The plan of this paper is as follows.
First, we define the model (Sec.\ \ref{model}),    
	and then 
	we derive the quantum effects 
	(Secs.\ \ref{fluctuations}--\ref{calculation}). 
We define the brane fluctuations (Sec.\ \ref{fluctuations}), 
	formulate the quantum effects (Sec.\ \ref{quantum}), 
	specify the method to regularize the divergences (Sec.\ \ref{cutoff}), 
	classify the possible induced terms according to symmetries
	 (Sec.\ \ref{classification}), 
	and calculate them via Feynman diagram method (Sec.\ \ref{calculation}).
Then, we interpret the quantum induced terms,
	and show why the induced gravity can avoid the problem
	(Secs.\ \ref{cosmological} and \ref{induced gravity}).
The cosmological terms are fine-tuned (Sec.\ \ref{cosmological}), and
	the Einstein like gravity and other terms are induced 
	(Sec.\ \ref{induced gravity}).
The final section (Sec.\ \ref{discussions}) is devoted to discussions.

\section{The Model \label{model}}

We consider a quantum theoretical braneworld described 
	by the  Dirac-Nambu-Goto type Lagrangian. 
We will see the quantum effects of the brane fluctuations
	give rise to effective braneworld gravity. 
Let $X^I(x^\mu)$ $(I=0,1,\cdots,D-1)$ 
	be the position of our three-brane in the $D$ dimensional spacetime (bulk), 
	parameterized by the brane coordinate $x^\mu$ $(\mu=0,1,2,3)$,
	where $I=0$ and $\mu=0$ indicate the time components.
Let $G^{IJ}(X^K)$ be the bulk metric tensor at the bulk point $X^K$.
This is taken to obey some bulk gravity theory.
Then we consider a braneworld with dynamics given 
	by the Dirac-Nambu-Goto type Lagrangian (density) \cite{DiracNambuGoto}:
\begin{eqnarray}
	{\cal L}_{\rm br}=-\lambda \sqrt{
	-\det_{\mu\nu} \left(
	\frac{\partial X^I}{\partial x _\mu}
	\frac{\partial X^J}{\partial x _\nu}
	G_{IJ}(X^K)
	\right)},
  \label{NG}
\end{eqnarray}
	where $\lambda$ is a constant.
Or we write it as
\begin{eqnarray}
	{\cal L}_{\rm br}=-\lambda \sqrt{- g^{[X]} }  \label{NGab}
\end{eqnarray}
with abbreviations $ g^{[X]} =\det g^{[X]}_{\mu\nu}$, and
\begin{eqnarray}
	{ g }^{[X]} _{\mu\nu} =X^I_{\ ,\mu} X^J_{\ ,\nu} G_{IJ}(X^K), 
  \label{gmunuX}
\end{eqnarray}
where (and hereafter) indices following a comma (,) indicate differentiation
	with respect to the corresponding coordinate component, 
	and ${[X]}$ is attached to remind 
	that they are abbreviations for expressions written in terms of $X^I$.
Note that $g^{[X]}_{\mu\nu}$ is the induced metric on the brane
	with (\ref{gmunuX}). 
We assume that $X^I$ appears nowhere other than in ${\cal L}_{\rm br}$ 
	in the total Lagrangian ${\cal L}_{\rm tot}$ including the bulk Lagrangian.
The system is invariant under the general coordinate transformation 
	of the bulk and the brane separately.
The Lagrangian (\ref{NG}) is the simplest among those with this symmetry.

Note that we do not a priori have the kinetic term of the metric
	in the basic Lagrangian (\ref{NG}).
It will be induced through quantum effects.
The metric emerges as a composite field, 
	and the gravitation is an induced phenomenon but not an elementary one.
If this is successful, we achieve a great conceptual advantage, 
	since it means that the familiar and important phenomenon of the gravitation
	is explained by the more fundamental ingredients.
Thus, it is worthwhile and urgent to examine 
	what is true and what is not in this idea.
This is the basic spirit of the induced gravity theory
	\cite{Sakharov}--\cite{indg3}.
As for the dynamics of the bulk metric,
	the situation is more vague and ambiguous.
We do not know what is the dynamics, i.e. what kinetic term we have, 
	and it is not clear even whether the kinetic term exists or not,
	i.e. whether it is dynamical or not.
Every case well deserves intensive investigations.
It is too naive to take that the bulk Einstein equation
	provides the brane Einstein equation at the brane,
	because large discrepancies take place 
	due to the extrinsic curvature terms.
The brane moves according to its equation of motion
	and is deformed in manners different from 
	what the naive gravity equation indicates. 
One may elaborate the elementary brane gravity from the elementary bulk gravity
	under specific conditions by hand in specific models 
	\cite{RS}, \cite{KantiKoganOlivePospelov}.
On the other hand,
	many people relied on the mechanism of brane induced gravity 
	to realize gravitation on the brane
	\cite{Akama82}, \cite{Pavsic},
	\cite{Akama87}--\cite{Akama88b},
	\cite{BIG}, \cite{DvaliGabadadze}, 
	\cite{AH}, \cite{Krause:2008sj}.
In the spirit of the brane induced gravity, 
	we do not a priori assume elementary gravity,
	and we take it to be induced from more fundamental ingredients.

Now, the equation of motion from (\ref{NG}) is given by
\begin{eqnarray}
	g ^{[X]} {}^{\mu\nu} X ^{[X]}_{;\mu\nu}{} ^I =0,
  \label{NGEM}
\end{eqnarray}
	where $ X ^{[X]}_{;\mu\nu} {}^I $ is the double covariant derivative
	with respect to both of the general coordinate transformations 
	on the brane and to those in the bulk:
\begin{eqnarray} 
	X ^{[X]}_{;\mu\nu} {}^I = X^I_{\ ,\mu\nu}
	-X^I_{\ ,\lambda} \gamma^{[X]}{}^\lambda_{\mu\nu}
	+X^J_{\ ,\mu} X^K_{\ ,\nu}\Gamma^I_{JK}
\label{X[X];munu}
\end{eqnarray}
	with the affine connections on the brane and bulk
\begin{eqnarray}
	\gamma^{[X]}{}^\lambda_{\mu\nu}
	&=&\frac{1}{2} g^{[X]}{} ^{\lambda \rho}\left(
	g ^{[X]}_{\rho\mu,\nu}+ g^{[X]}_{\rho\nu,\mu}- g^{[X]} _{\mu\nu,\rho}
	\right),
\\
	\Gamma^I_{JK}
	&=&\frac{1}{2} G^{IL}\left(
	G_{LJ,K}+ G_{LK,J}- G_{JK,L}
	\right),
\label{GammaBulk}
\end{eqnarray}
	respectively.
We expect that this gives a good approximation at low curvature limit 
	in many dynamical models of the braneworld 
	(e.g. topological defects, spacetime singularities, D-branes, etc.).
It is remarkable that, as we shall see below, this simple model exhibits 
	brane gravity and gauge theory like structure through the quantum effects.

For convenience of quantum treatments, 
	we consider the following equivalent Lagrangian 
	to (\ref{NG}):
\begin{eqnarray}
	{\cal L}'_{\rm br}=-\frac{\lambda}{2} \sqrt{-g}\left[
	g^{\mu\nu}X^I_{\ ,\mu} X^J_{\ ,\nu}G_{IJ}(X^K)-2
	\right]
  \label{NG'}
\end{eqnarray}
where $g_{\mu\nu}$ is an auxiliary field, $g=\det g_{\mu\nu}$, and
	$g^{\mu\nu}$ is the inverse matrix of $g_{\mu\nu}$.
Note that $g_{\mu\nu}$, unlike $ g^{[X]}_{\mu\nu}$ above,
	is treated as a field independent of $X^I$.
Then the Euler Lagrange equations with respect to $X^I$ and $g_{\mu\nu}$ 
	are given by 
\begin{eqnarray}&&\hskip-20pt
	g^{\mu\nu} X _{;\mu\nu} ^I =0,
  \label{NG'EM}
\\&&\hskip-20pt 
	g_{\mu\nu}=X^I_{\ ,\mu} X^J_{\ ,\nu}G_{IJ}(X^K),
  \label{gmunu}
\end{eqnarray}
	respectively, where the covariant derivative 
\begin{eqnarray}
	X _{;\mu\nu}{} ^I = X^I_{\ ,\mu\nu}
	-X^I_{\ ,\lambda} \gamma^\lambda_{\ \mu\nu}
	+X^J_{\ ,\mu} X^K_{\ ,\nu}\Gamma^I_{\ JK}
\label{X;munu}
\end{eqnarray}
is written in terms of the brane affine connection 
\begin{eqnarray}
	\gamma^\lambda_{\ \mu\nu}
	&=&\frac{1}{2} g ^{\lambda \rho}\left(
	g _{\rho\mu,\nu}+ g _{\rho\nu,\mu}- g _{\mu\nu,\rho}
	\right)
\label{gamma}
\end{eqnarray}
	with respect to the auxiliary field $g_{\mu\nu}$.
Now $g_{\mu\nu}$ in (\ref{NG'EM}) is independent of $X^I$,
	and, instead, we have an extra equation (\ref{gmunu}),
	which guarantees that $g_{\mu\nu}$ is the induced metric.
If we substitute (\ref{gmunu}) into (\ref{NG'EM}), 
	we obtain the same equation as (\ref{NGEM}).
Thus the systems with the Lagrangians ${\cal L}_{\rm br}$ 
	and ${\cal L}'_{\rm br}$ coincide.
Furthermore the argument that their Dirac bracket algebrae coincide 
	\cite{Akama79} indicates their quantum theoretical equivalence.
We proceed hereafter based on the Lagrangian ${\cal L}'_{\rm br}$
	instead of ${\cal L}_{\rm br}$.

\section{Brane Fluctuations\label{fluctuations}}

In order to extract the quantum effects of ${\cal L}'_{\rm br}$,
	we deploy a semi-classical method, 
	where we consider those due to small fluctuations of the brane 
	around some classical solution (say $Y^I(x^\mu)$) for $X^I(x^\mu)$
	of the equation of motion (\ref{NGEM}) \cite{Watanabe}.
Namely, the solution $Y^I(x^\mu)$ obeys the classical equation
\begin{eqnarray}&&
	g^{\mu\nu} Y _{;\mu\nu}^I =0,
  \label{NG'EMY}
\\&&
	g_{\mu\nu}=Y^I_{\ ,\mu} Y^J_{\ ,\nu}G_{IJ}(Y^K).
  \label{gmunuY}
\end{eqnarray}
In quantum treatment, 
	$X^I$ itself in the Lagrangian ${\cal L}'_{\rm br}$ does not necessarily
	obey the equation of motion (\ref{NGEM}), 
	and may fluctuate from $Y^I(x^\mu)$. 
{\it Among the fluctuations, 
	only those transverse to the brane are physically meaningful},
	because those along the brane remain within the brane
	and cause no real fluctuations of the brane.   
They are absorbed by general coordinate transformations.
In order to describe them, we choose $D-4$ independent normal vectors 
	$n_m{}^I (x^\mu)$ $(m=4,\cdots,D-1)$ 
	at each point on the brane with the normality condition
\begin{eqnarray}&&
	n_m{}^I Y^J_{\ ,\nu} G _{IJ}(Y^K) =0.
  \label{nY}
\end{eqnarray}
Then we express the fluctuations as
\begin{eqnarray}
	X^I(x^\mu)= Y^I(x^\mu)+ \phi^m(x^\mu)n_m{}^{I} (x^\mu),
  \label{fluctuation}
\end{eqnarray}
	where $\phi^m(x^\mu)$ is the transverse fluctuation 
	along $n_m{}^{I} (x^\mu)$ $(m=4,\cdots,D-1)$.
The arbitrariness of $n_m{}^{I}$ in choice under the condition (\ref{nY}) 
	gives rise to gauge symmetry under the group GL($D-4$) of 
	the general linear transformations of the normal space
	at each point on the brane.
If we define $g_{mn} = n_m^{\ I} n_n^{\ J } G_{IJ}(Y^K)$ and its inverse $g^{mn}$, 
	we have the completeness condition, 
\begin{eqnarray}&&
	Y^I_{\ ,\mu}Y^J_{\ ,\nu} g^{\mu\nu}+n_m^{\ I} n_n^{\ J} g^{mn}
	= G ^{IJ}(Y^K).
  \label{YY+nn=G}                                                        
\end{eqnarray}
Bulk coordinate indices $I,J,\cdots(=0,\cdots,D-1) $ are raised and lowered
	by the metric tensors $G_{IJ}$ and $G^{IJ}$.
We can read off from (\ref{NG'}) and (\ref{gmunu})
	that the auxiliary field $g_{\mu\nu}$ 
	plays the role the metric tensor on the brane.
Hereafter we raise and lower 
	the brane coordinate indices $\mu,\nu,\cdots(=0,\cdots,3) $ 
	by $g_{\mu\nu}$ and $g^{\mu\nu}$ 
	(but not by $g^{[X]}_{\mu\nu}$ and $g^{[X] \mu\nu}$).
We raise and lower 
	the normal space indices $m,n,\cdots(=4,\cdots,D-1) $ 
	by  $g_{mn}$ and $g^{mn}$.

A problem of the definition (\ref{fluctuation}) of $\phi^m$
	is that $\phi^m$ lacks the bulk general-coordinate invariance.
In fact, it is transformed in a complex way  
	under the general coordinate transformations of the bulk.
On the contrary, the invariant definition of $\phi^m$
	requires a complex formula instead of (\ref{fluctuation}).
For quantum treatments, however, it is desirable to have a relation linear 
	in $\phi^m$ like (\ref{fluctuation}).
It requires further careful considerations.
Therefore, in this paper, we restrict ourselves to the case where the bulk is flat.
Namely, there exists the cartesian frame with
\begin{eqnarray}
	G_{IJ}=\eta_{IJ} \equiv {\rm diag}\{1,-1,-1,\cdots,-1\},
\end{eqnarray}
where we have $\Gamma^I_{JK}=0$. 
In this case, we have $R^I{}_{JKL}=0$ in any frame,
	and $\phi^m$ restores the bulk general-coordinate invariance.
The general case of curved bulk will be considered in a separate 
	forthcoming paper.

Now we substitute (\ref{fluctuation}) into the Lagrangian (\ref{NG'}), and obtain
\begin{eqnarray}&&\hskip-15pt
	{\cal L}'_{\rm br}= {\cal L}'_0 + {\cal L}'_{\phi}\ \ \ \ \ {\rm with}
\\&&\hskip-15pt	
	{\cal L}'_0= {\cal L}'_{\rm br}|_{\phi=0}=-\frac{\lambda}{2} \sqrt{-g} \Big[
	g^{\mu\nu}Y^I_{\ ,\mu} Y^J_{\ ,\nu}G _{IJ}-2\Big], 
\\&&\hskip-15pt	
	{\cal L}'_{\phi}=-\frac{\lambda}{2} \sqrt{-g} \Big[
	g^{\mu\nu}g^{ \rho \sigma}
	\phi^m\phi^n B_{m\mu\rho }B_{n\nu \sigma} 
\cr&&\hskip-10pt	
	+ g^{\mu\nu}g_{mn}
	(\phi^m_{,\mu}+ g^{mi} \phi^k A_{ik\mu})
	(\phi^n_{,\nu}+ g^{nj} \phi^l A_{jl\nu})
	\Big], 
  \label{S'1}
\end{eqnarray}
where $A_{mn\mu}$ and $B_{m\mu\nu}$ are the normal connection
	and the extrinsic curvature, respectively. 
They are given by  
\begin{eqnarray}
	A_{mn\mu}&=&n_m^{\ I}n_{n;\mu}^{\ J}G_{IJ}, 
  \label{Amnmu}
\\
	B_{m\mu\nu}&=&n_m^{\ I} Y _{;\mu\nu}^J G _{IJ}, 
  \label{Bmmunu}
\end{eqnarray}
where $n_ {n;\mu }^J $ is the covariant derivative:
\begin{eqnarray}
	n_ {n;\mu }^J = n^{ \ J}_{n\ ,\mu} 
	+ n_n^K Y^M_{,\mu}\Gamma^J_{KM}, 
  \label{Dn}
\end{eqnarray}
which coincides with the ordinary derivative $ n^{ \ J}_{n\ ,\mu} $
	under the assumption $G_{IJ}=\eta_{IJ}$ for the present paper.
We can see that (\ref{S'1}) is the Lagrangian 
	for the quantum scalar fields $\phi^m$ on the curved brane 
	interacting with the given external fields 
	$A_{mn\mu}$ and $B_{m\mu\nu}$.
For later use, it is convenient to rewrite (\ref{S'1}) into 
\begin{eqnarray}&&\hskip-15pt
	{\cal L}'_{\phi}=-\frac{\lambda}{2} \sqrt{-g} \phi^m \Big[ 
	-\partial_\mu \sqrt{-g} g ^{\mu\nu} \partial_\nu 
\cr&&	-\partial_\mu\sqrt{-g} A^m{}_n{}^\mu 
	-\sqrt{-g} A^m{}_n{}^\mu \partial_\mu 
\cr&&
	-             \sqrt{-g} (A^{mk\mu} A_{kn\mu}-B^{m\mu\rho }B_{n\mu\rho })
	             \Big]\phi^n,
  \label{L'phi}
\end{eqnarray}
where total derivatives are neglected.

\section{Quantum Effects\label{quantum}}

\begin{figure}
\includegraphics[width=6.0cm]{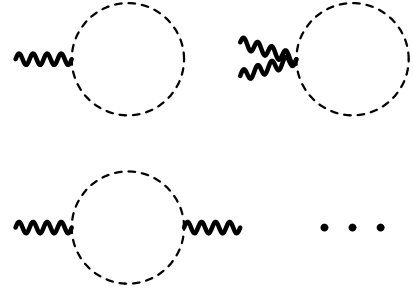}
\caption{The Feynman diagrams.
The dashed lines indicate the $\phi^m$-propagators,
	and the wavy lines indicate external fields 
	of $A_{mn\mu}$, $B_{m\mu\nu}$ or $h_{\mu\nu}$. 
The dots indicate an infinite series of diagrams 
with a dashed-line loop and with more external wavy lines than two.
By virtue of the symmetries of the system, we have only to calculate
	the three diagrams explicitly drawn here.
}
\label{fig1}
\end{figure}

The quantum effects of the field $\phi^m$ are described 
	by the effective Lagrangian $ {\cal L}^ {\rm eff }$ 
\begin{eqnarray}&&\hskip-15pt
	\int {\cal L}^{\rm eff}d^4x
	=-i \ln\int [d\phi^m] \exp \left [i\int {\cal L}'_{\phi} d^4x\right],
  \label{Seff}
\end{eqnarray}
	where $[d\phi^m] $ is the path-integration over $\phi^m $. 
To perform it, we rewrite (\ref{L'phi}) into the form
\begin{eqnarray}
	{\cal L}'_{\phi}
	= \frac{\lambda}{2} \phi_m \left(
	-\delta^m_n \Box + {\cal V}^m{}_n 
	\right) \phi^n,
  \label{S'V}
\end{eqnarray}
	with $\Box=\eta^{\mu\nu}\partial_\mu\partial_\nu $ and
\begin{eqnarray}
	{\cal V} ^m{}_n
	=\delta^m_n \partial_{\mu} {\cal H} ^{\mu\nu} \partial_{\nu}
	+\partial_{\mu}  {\cal A}^m{}_n{}^\mu 
	+{\cal A}^m{}_n{}^\mu \partial_{\mu}
	+ {\cal Z}^m{}_n
  \label{Vmn}
\end{eqnarray}
\begin{eqnarray}
	{\cal H} ^{\mu\nu}\ \ 
	&=& \eta^{\mu\nu} -\sqrt{-g} g ^{\mu\nu}
  \label{calH=}
\\	
	{\cal A}^m{}_n{}^\mu &=& -\sqrt{-g} A^m{}_n{}^\mu 
  \label{calA=}
\\	
	{\cal Z}^m{}_n{}\hskip5pt &=&
	-\sqrt{-g} (A^{mk\mu} A_{kn\mu}-B^{m\mu\rho }B_{n\mu\rho }),
  \label{calZ=}
\end{eqnarray}
where the differential operator
	$\partial_\mu\equiv\partial/\partial x ^\mu $ 
	is taken to operate on the whole expression 
	in its right side in (\ref{S'V}). 
The path-integration in (\ref{Seff}) is performed to give 
\begin{eqnarray}
	\int {\cal L}^{\rm eff} d^4 x
	&=&\sum_{l=0}^{\infty}\frac{1}{2li}{\rm Tr}
	\left(\frac{1}{\Box} {\cal V}^m{}_n \right)^l,
  \label{S'eff1}
\end{eqnarray}
	up to additional constants,
	where Tr indicates the trace over the brane coordinate variable $x^\mu$
	and extra dimension index $m$.
The terms in (\ref{S'eff1}) can be calculated with Feynman-diagram method. 
In terms of the Fourier transforms
\begin{eqnarray}
	\tilde {\cal H}^{\mu\nu}(q_i)
	&=&\int d^4x{\cal H}^{\mu\nu}(x)
	e^{iq_i x} ,
  \label{FourierH}
\\
	\tilde {\cal A}^m{}_n{}^\mu (q_i)
	&=&\int d^4x{\cal A}^m{}_n{}^\mu (x)
	e^{iq_i x} ,
  \label{FourierA}%
\\
	\tilde {\cal Z}^m{}_n(q_i)
	&=&\int d^4x{\cal Z}^m{}_n(x)
	e^{iq_i x} ,
  \label{FourierV}%
\end{eqnarray}
the effective Lagrangian ${\cal L}^{\rm eff}$ is written as
\begin{eqnarray}&&
	{\cal L}^{\rm eff}= \sum_{l=0}^{\infty}\frac{1}{2l}
	\prod_{i=1}^{l}\int\frac{d^4q_i}{(2\pi)^4}e^{-iq_i x} 
	G^l,
  \label{FourierGl}%
\\&&
	G^l
	=\int\frac{d^4p}{i(2\pi)^4}
	\prod_{i=1}^{l}\frac{1}{-p_i^2}
	\tilde {\cal V}^{m_i}{}_{m_{i-1}}(p_i,q_i),
  \label{Gl=}%
\\ &&
	\tilde {\cal V}^m{}_n (p_i,q_i)
	=-\delta^m_n
	(p_i{})_\mu( p_{i-1}){}_\nu \tilde{\cal H} ^{\mu\nu}(q_i) 
\cr&&\ \ \ \ \ \ \ 
	-i (p_i+ p_{i-1})_\mu  \tilde {\cal A}^m{}_n{}^\mu (q_i)
	+ \tilde {\cal Z}^m{}_n(q_i),
  \label{tildeVmn}
\end{eqnarray}
where $p_i=p+q_1+\cdots+q_{i}$ and $m_0=m_l$. 
The function $G^l$ is nothing but the Feynman amplitude for
	the one-loop diagram with $l$ internal lines of $\phi^m$ and 
	$l$ vertices of $\tilde {\cal V}^m{}_n $ (FIG.\ \ref{fig1}).
Unfortunately, the $p$-dependence of the integrand 
	in (\ref{FourierGl}) with (\ref{Gl=}) indicates that 
	the integration over $p$ diverges at most quartically.
The divergences will be regulated in the next section.
Then, we can perform the integration over $p$ 
	to obtain the function $G^l$. 
The $q_k$'s are replaced by differentiation $i\partial_k$
	of the $k$-th vertex function 
	according to the inverse Fourier transformation in (\ref{FourierGl}).
Collecting all the contributions, 
	which are functions of the fields 
	$g_{\mu\nu}$, $A_{mn\mu}$ and $B_{m\mu\nu}$ and their derivatives,
	we can obtain the expression for 
	the effective Lagrangian ${\cal L}^{\rm eff}$.

\section{Divergences and Regularization \label{cutoff}}

The $p$-dependence of the integrand in (\ref{FourierGl}) with (\ref{Gl=})
	indicates that 
	the integration over $p$ diverges at most quartically.
We expect, however, that fluctuations with smaller wave length
	than the brane thickness are suppressed.
Then, the momenta higher than the inverse of the thickness are cut off.
In order to model the cutoff without violating full symmetry of ${\cal L}'_{\rm br}$,
	we introduce three Pauli-Villers regulators $\Phi_j^m$
	with very large mass $M_j$ ($j$=1,2,3), 
	which are taken equal finally, $M_j\rightarrow\Lambda$,
	following the original paper \cite{Akama78}.
Precisely, it amounts to consider the regularized effective Lagrangian 
\begin{eqnarray}
	{\cal L}^{\rm reg}={\cal L}^{\rm eff}+\sum_{j=1}^3 C_j {\cal L}^{\rm eff}_{M_j} 
  \label{Sreg}
\end{eqnarray}
where $C_j$ are the coefficients defined by
\begin{eqnarray}
	\sum_{j=1}^3 C_j (M_j)^{2k}=-\delta^{0k}\ (k=0,1,2),
  \label{sumCM2k}
\end{eqnarray}
and ${\cal L}^{\rm eff}_{M_j}$ is the effective Lagrangian 
	for the quantum effects from ${\cal L}'_{\Phi_j} $ 
	which is the same as ${\cal L}'_{\phi}$ except that
	$\phi^m$ is replaced by the regulator field $\Phi_j^m$ with mass $M_j$.
\begin{eqnarray}&&\hskip-15pt
	\int {\cal L}^{\rm eff}_{M_j}d^4 x
	=-i \ln\int [d\Phi_j^m] \exp \left[i\int {\cal L}'_{\Phi_j} d^4 x\right].
  \label{SeffMj}
\\&&\hskip-15pt
	{\cal L}'_{\Phi_j}= {\cal L}'_{\phi}|_{\phi=\Phi_j}+
	\frac{ 1}{2}\lambda M_j^2 \sqrt{-g} \Phi_j^m\Phi_j^n \eta_{mn}.
  \label{S'Mj}
\end{eqnarray}
Note that the added mass term also preserves the full symmetry of ${\cal L}'_{\rm br}$. 
Performing the path integration over $\Phi_j^m $, we have
\begin{eqnarray}&&
	\int {\cal L}^{\rm eff}_{M_j} d^4 x
	=\sum_{l=0}^{\infty}\frac{1}{2li}{\rm Tr}
	\left(\frac{1}{\Box+M_j^2}{\cal V}_{M_j}{}^m{}_n \right)^l,
  \label{Seffj}
\\&&
	{\cal V}_{M_j}{}^m{}_n ={\cal V}^m{}_n+{\cal F} M_j^2\delta^m_n,
  \label{VMjmn}
\\&&
	{\cal F} =1-\sqrt{-g},
  \label{calF=}
\end{eqnarray}
with ${\cal V}^m{}_n $ in (\ref{Vmn}).
In terms of the Fourier transform 
\begin{eqnarray}&&
	\tilde {\cal F} (q_i) 	=\int d^4x{\cal F} (x) 	e^{iq_i x} .
  \label{FourierF}%
\end{eqnarray}
we have
\begin{eqnarray}&&
	{\cal L}_{M_j}^{\rm eff}= \sum_{l=0}^{\infty}\frac{1}{2l}
	\prod_{i=1}^{l}\int\frac{d^4q_i}{(2\pi)^4}e^{-iq_i x} 
	G_{M_j}^l,
  \label{FourierGMjl}%
\\&&
	G_{M_j}^l
	=\int\frac{d^4p}{i(2\pi)^4}
	\prod_{i=1}^{l}\frac{1}{-p_i^2+M_j^2}
	\tilde {\cal V}_{M_j}{}^{m_i}{}_{m_{i-1}}(p_i,q_i),
  \label{GMjl=}%
\cr&&
\\ && 
	\tilde {\cal V}_{M_j}{}^m{}_n (p_i,q_i)
	=\tilde {\cal V}{}^m{}_n (p_i,q_i)+\delta^m_n M_j^2\tilde{\cal F}(q_i), 
  \label{tildeVMjmn}
\end{eqnarray}
with $\tilde {\cal V}{}^m{}_n (p_i,q_i)$ in (\ref{tildeVmn}).
In dimensional regularization, 
	the divergent parts of the Feynman amplitude $G_{M_j}^l$ behaves like
\begin{eqnarray}&&\hskip-15pt
	G_{M_j}^l
	\sim \epsilon^{-1}
	({\cal G}_4M_j^4+{\cal G}_2M_j^2+{\cal G}_0),
  \label{GMjleps}
\end{eqnarray}
	where this is evaluated at the spacetime dimension $4-2\epsilon$,
	and ${\cal G}_{2k}$ are the appropriate coefficient functions.
The singularities at $\epsilon=0$ reflect the divergences in the $p$-integration.
We can see that, 
	when they are summed with the coefficients $C_j$ over $j$ in (\ref{Sreg}),
	they cancel out according to (\ref{sumCM2k}).
Therefore the $p$-integrations in ${\cal L}^{\rm reg}$ converge.
Any positive power contributions of $M_j$ regular at infinity
	vanish according to (\ref{sumCM2k}).
The function $G_{M_j}^l$ involves logarithmic singularities in $M_j$, 
	which tend, in the equal mass limit $M_j\rightarrow \Lambda$, 
\begin{eqnarray}&&
	\sum_{j=1}^3 C_j M_j^4\ln M_j^{2}\rightarrow -\Lambda^4/2
  \label{CM4ln->}
\\&&
	\sum_{j=1}^3 C_j M_j^2\ln M_j^{2}\rightarrow \Lambda^2/2
  \label{CM2ln->}
\\&&
	\sum_{j=1}^3 C_j \ln M_j^{2}\rightarrow -\ln\Lambda^2
  \label{Cln->}
\end{eqnarray}

\section{Classification of the Terms \label{classification}}

Thus the divergent part ${\cal L}^{\rm div}$ of the regularized effective
	Lagrangian ${\cal L}^{\rm reg}$ consist of the terms 
	which are proportional to $\Lambda^4$, $\Lambda^2$ or $\ln\Lambda^2$,
	and are monomials of ${\cal H}^{\mu\nu}$, ${\cal F}$,
	${\cal A}^m{}_n{}^\mu $, ${\cal Z}^m{}_n$, and their derivatives.
The expressions ${\cal H}^{\mu\nu}$, ${\cal F}$,
	${\cal A}^m{}_n{}^\mu $, and ${\cal Z}^m{}_n$ 
	are written in terms of the fields 
	$g_{\mu\nu}$, $A^m{}_n{}^\mu$, and $B^m{}_{\mu\nu}$
	according to (\ref{calH=}),  (\ref{calF=}), (\ref{calA=}), 
	and (\ref{calZ=}).
Introducing the notation 
	$h_{\mu\nu}\equiv g_{\mu\nu}-\eta_{\mu\nu}$,
	we rewrite $ g^{\mu\nu}$ and $\sqrt{-g}$ in ${\cal H}^{\mu\nu}$, ${\cal F}$,
	${\cal A}^m{}_n{}^\mu $, and ${\cal Z}^m{}_n$
	according to
\begin{eqnarray}
	g^{\mu\nu}&=& \eta^{\mu\nu}-h^{\mu\nu}+h_{(2)}^{\mu\nu} 
	+h_{(3)}^{\mu\nu} +\cdots ,
  \label{g^mumu-inh}
\\
	\sqrt{-g}&=& 1+h/2- h_{(2)}/4+ h^2/8+\cdots, 
  \label{sqrt-g-inh}
\end{eqnarray}
with \cite{h^superscripts}
\begin{eqnarray}&&
	h_{(n)}{}^\mu{}_\nu
	=\overbrace{ h^\mu{}_\sigma h^\sigma{} _\tau \cdots h^\rho{} _\nu } ^n,
  \label{h(2)munu}
\\&&
	h = h^\mu{} _\mu,\ \ \ \ \ \ 
	h_{(n)}= h_{(n)}{} ^\mu{} _\mu. 
  \label{h,h(2)}
\end{eqnarray}
Then, ${\cal L}^{\rm div}$ becomes an infinite sum of monomials of 
	$h_{\mu\nu}$, $A^m{}_n{}_\mu$, $B^m{}_{\mu\nu}$,
	and their derivatives.
Let us denote the numbers of $h_{\mu\nu}$, 
	$A^m{}_{n\mu}$, $B^m{}_{\mu\nu} $, and the differential operators
	in the monomial by $N_h$, 
	$N_A$, $N_B$ and $N_{\partial}$, respectively. 
The Lagrangian ${\cal L}^{\rm reg}$ should have mass dimension 4,
	while $h_{\mu\nu}$, 
	$A^m{}_{n\mu}$, $B^m{}_{\mu\nu} $, and the differential operator
	has mass dimension 0, 1, 1, and 1, respectively.
Therefore, the numbers $N_A$, $N_B$ and $N_{\partial}$ are restricted by
\begin{eqnarray}
	N_A+ N_B+ N_{\partial}\le 4-2k_{\rm div},
  \label{N+N+N<}
\end{eqnarray}
where $ k_{\rm div}=2,1,0$ for $\Lambda^4$, $\Lambda^2$, and $\ln\Lambda^2$
	terms, respectively. 
On the other hand, the number $N_h$ of $h_{\mu\nu}$ is not restricted.
The relation (\ref{N+N+N<}) allows only finite numbers of values of
	$N_A$, $N_B$ and $N_{\partial}$,
	according to which we can classify the terms of ${\cal L}^{\rm div}$.
Each class involves infinitely many terms for arbitrary values of $N_h$.

They are, however, not all independent, because they are related by 
	high symmetry of the system
	under the general coordinate transformations on the brane
	and GL(4) gauge transformations of the normal space rotation.
Though the original system is invariant under the  
	general coordinate transformations in the bulk also,
	it is not available here, 
	because we restrict ourselves to the flat bulk in this paper.
The general case will be discussed in the forthcoming paper.
Owing to the symmetry of the system,
	only finite number of terms are allowed. 
The general coordinate transformation symmetry requires that 
	the effective Lagrangian density is proportional to $\sqrt{-g}$
	times a sum of invariant forms.
We list the allowed invariant forms in TABLE \ref{tab1}.
\begin{table}
\caption{\label{tab1}Invariant forms}
\begin{ruledtabular}
\begin{tabular}{ccccl}
$k^{\rm div}$&$N_A$&$N_B$&$N_{\partial}$&invariant forms \\
\hline
2&0&0&0&1\\
\hline
1&0&0&2&$R$\\
 &0&2&0&$B^{(2)}$, $B_mB^m $\\
\hline
0&0&0&4&
$R^2$, $R_{\mu\nu} R^{\mu\nu}$, $R_{\mu\nu\rho\sigma} R^{\mu\nu\rho\sigma }$\\
&0&2&2&
$RB^{(2)}$, $RB_mB^m$, $R_{\mu\nu}B^{m\mu\nu}B_m$,\\
&&&&
$R_{\mu\nu}B_{m\lambda}{}^{\mu}B^{m\nu\lambda}$,
$R_{\mu\nu\rho\sigma} B_{m\lambda}{}^{\mu}B^{m\nu\lambda}$\\
&0&4&0&
$(B^{(2)})^2$, $(B_mB^m)^2$, $ B^{(2)} B_mB^m$,\\
&&&&
$ B^{(4)} $,\ \ 
$B_{m\mu\lambda}B^{m\mu\rho} B_{n\nu\rho}B^{n\nu\lambda}$\\
&&&&
$B^mB_{m\mu\nu}B^{n\mu\rho} B_{n\rho}{} ^{ \nu }$\\
&&&&
$B^mB_nB_{m\mu\nu}B^{n\mu\nu}$\\
&0,1,2&2&$2-N_A$&
$B_{m\mu\nu\|\lambda} B^{m\mu\nu\|\lambda }$,\ \ 
$B_{m\mu\nu}{}^{\|\nu} B^{m\mu\lambda}{}_{\|\lambda }$\\
&&&&
$B_{m\mu\nu\|\lambda} B^{m \|\mu}$,\ \ 
$B_{m\|\mu} B^{m\|\mu}$,\\
&1,2&2&$2-N_A$&
$ A_{mn\mu\nu}B^{m\mu\lambda} B^{n\nu }{}_{\lambda }$\\
&2,3,4&0&$4-N_A$&
$A_{mn\mu\nu} A^{mn\mu\nu}$\\
\end{tabular}
\end{ruledtabular}
\end{table}
In the table and thereafter, we use the following abbreviations.
\begin{eqnarray}&&
	B^{(2)}= B^{m\mu\nu}B_{m\mu\nu},  \ \ \ 
	B_m=B_{m\nu}{}^{\nu}.  
  \label{BB}
\\&&
	B^{(4)}= B_{m\mu\nu }B^{m \rho\lambda } B_{n \rho\lambda }B^{n\mu \nu }.  
  \label{BBBB}
\\&&
	B_{m\|\lambda}= B_{m;\lambda}
	+A_{m}{}^{n}{}_{\lambda} B_{n }.  
  \label{Bm;}
\\&&
	B_{m\mu\nu\|\lambda}= B_{m\mu\nu;\lambda}
	+A_{m}{}^{n}{}_{\lambda} B_{n\mu\nu}.  
  \label{Bmmunu;}
\\&&
	A_{mn\mu\nu}=
	\partial_{ \mu} A_{mn \nu}-\partial_{ \nu} A_{mn \mu}
\cr&&\hskip40pt	
	+A_{mk \mu} A^k_{\ n \nu}-A_{mk \nu } A^k_{\ n\mu },
  \label{Amnmunu}
\end{eqnarray}
where (\ref{Bm;}) and (\ref{Bmmunu;}) are 
	the covariant derivatives of the full symmetry,
and (\ref{Amnmunu}) is the field strength of the gauge field $A_{m\mu\nu}$.
In the table, $R^2$, $R_{\mu\nu} R^{\mu\nu}$, 
	and $R_{\mu\nu\rho\sigma} R^{\mu\nu\rho\sigma }$
	are not all independent, but related by Gauss-Bonnet relation,
	and some other combinations are related due to
	the Gauss-Codazzi-Ricci formulae.

\section{Calculation \label{calculation}}

Thus we can calculate the coefficients 
	of the term $\sqrt{-g}$ times the invariant forms
	by calculating the lowest order contributions in $h_{\mu\nu}$.
The lowest contributions to the term with $N_A=N_B=N_\partial=0$ 
	are $O(h_{\mu\nu})$,
	while those to $N_A=N_B=0$ and  $ N_\partial\not=0$ 
	are $O((h_{\mu\nu})^2)$,
	because the  $O(h_{\mu\nu})$ terms are total derivatives.
Therefore, their lowest terms are in the one- and two-point functions $G^1$ and $G^2$.
We can see from (\ref{Vmn}) -- (\ref{calZ=}) that $B_{m\mu\nu}$ always appears 
	in the combination $B_{m\mu\nu}B^{n\mu\nu}$.
Therefore, the only possible forms including $B_{m\mu\nu}$ are
	$B^{(2)}$, $RB^{(2)}$, $(B^{(2)})^2$, and $B^{(4)}$,
	among many forms listed in Table \ref{tab1}.
Their lowest terms are those with $O((h_{\mu\nu})^0)$ 
	and are also in $G^1$ and $G^2$.
The only possible form including $A_{mn\mu}$ only is $ A_{mn\mu\nu} A^{mn\mu\nu}$
	and its lowest term is of $O((h_{\mu\nu})^0)$, and it is in $G^2$.
Thus, it suffices to calculate $G^1$ and $G^2$ 
	in order to determine full contributions to ${\cal L}^{\rm div}$.

From (\ref{GMjl=}), (\ref{tildeVMjmn}) and (\ref{tildeVmn}), 
	they are given by
\begin{eqnarray}&&\hskip-15pt
	G^1_{M_j}=-N_{\rm ex}\tilde {\cal H}^{\mu\nu}I_{\mu\nu}
	+N_{\rm ex}M_j^2\tilde {\cal F}I+ \tilde {\cal Z}^m{}_m I,
  \label{G^1=}
\\&&\hskip-15pt
	G^2_{M_j}=N_{\rm ex}
	\tilde {\cal H}^{\mu\nu} 
	\tilde {\cal H}^{\lambda\rho} J_{\mu\nu\lambda\rho }
	+2i\tilde {\cal H}^{\mu\nu}\tilde{\cal A}^n{}_n{}^\rho J_{\mu\nu\rho}
\cr&&
	-(N_{\rm ex}M_j^2\tilde {\cal F}+\tilde {\cal Z}^m{}_m)
	\tilde {\cal H}^{\mu\nu}(J_{\mu\nu}-q_\mu q_\nu J)/4
\cr&&
	- \tilde {\cal A}^m{}_n{}^\mu\tilde{\cal A}^n{}_m{}^\nu J_{\mu\nu}
	-2i(M_j^2\tilde {\cal F}\delta^m_n+{\cal Z}^m{}_n)
	\tilde{\cal A}^n{}_m{}^\rho J_{\mu}
\cr&&
	+ (N_{\rm ex}M_j^4\tilde {\cal F}\tilde {\cal F}
	+ M_j^2\tilde {\cal F}{\cal Z}^m{}_m
	+\tilde {\cal Z}^m{}_n \tilde {\cal Z}^n{}_m)J,
  \label{G^2=}
\end{eqnarray}
	where $ N_{\rm ex}=D-4$ is the number of the extra dimensions,
	$q_\mu$ is the momentum flowing in and out through the vertices,
	and
\begin{eqnarray}&&
	I_{\mu\nu}=\int \frac{d^4p}{i(2\pi)^4}
	\frac{p_\mu p_\nu}{[-p^2+M_j^2]},
\\&&
	I=\int \frac{d^4p}{i(2\pi)^4}
	\frac{1}{[-p^2+M_j^2]},
\\&&\hskip-20pt
	J_{\mu\nu\lambda\rho}=\int \frac{d^4p}{i(2\pi)^4}
	\frac{(p+q)_\mu p_\nu p_\lambda (p+q)_\rho }
	{[-(p+q)^2+M_j^2][-p^2+M_j^2]}.
\\&&\hskip-20pt
	J_{\mu\nu\rho}=\int \frac{d^4p}{i(2\pi)^4}
	\frac{(p+q)_\mu p_\nu (2p+q)_\rho }
	{[-(p+q)^2+M_j^2][-p^2+M_j^2]},
\\&&\hskip-20pt
	J_{\mu\nu}=\int \frac{d^4p}{i(2\pi)^4}
	\frac{(2p+q)_\mu (2p+q)_\nu}
	{[-(p+q)^2+M_j^2][-p^2+M_j^2]},
\\&&\hskip-20pt
	J_{\mu}=\int \frac{d^4p}{i(2\pi)^4}
	\frac{(2p+q)_\mu }
	{[-(p+q)^2+M_j^2][-p^2+M_j^2]},
\\&&\hskip-20pt
	J=\int \frac{d^4p}{i(2\pi)^4}
	\frac{1}
	{[-(p+q)^2+M_j^2][-p^2+M_j^2]},
\end{eqnarray}
In the dimensional regularization, for large $M_j^2$, they are calculated to be
\begin{eqnarray}&&\hskip-20pt 
	I_{\mu\nu}=- {\cal I}_j M_j^2\eta_{\mu\nu},\ \ \ I={\cal I}_j M_j^2,\ \ \ 
  \label{Iresult}
\\&&\hskip-20pt
	J_{\mu\nu\lambda\rho}= {\cal I}_j \Bigg[
	\left(\frac{M_j^4}{8}-\frac{M_j^2q^2}{24}+\frac{q^4}{240}\right)
	S_{\mu\nu\lambda\rho}
\cr&&
	-\left(\frac{M_j^2}{12}-\frac{q^2}{60}\right)
	T_{\mu\nu\lambda\rho}
	+\left(\frac{M_j^2}{6}-\frac{q^2}{40}\right)
	T'_{\mu\nu\lambda\rho}
\cr&& \hskip70pt 
	+\frac{1}{30}q_\mu q_\nu q_\lambda q_\rho 
	\Bigg], 
  \label{Jresult}
\\&&\hskip-20pt 
	J_{\mu\nu}
	=- {\cal I}_j [2M_j^2\eta_{\mu\nu}+( q_\mu q_\nu -q^2\eta _{\mu\nu})/3], 
\\&&\hskip-20pt 
	J_{\mu\nu\rho}=0,\ \ \ \  
	J_{\mu}=0, \ \ \ \  
	J=  {\cal I}_j 
\end{eqnarray}
with ${\cal I}_j={M_j^{-2\epsilon}}/(4\pi)^2\epsilon$ and
\begin{eqnarray}&&\hskip-25pt 
	S_{\mu\nu\lambda\rho}
	=\eta_{\mu\nu}\eta_{\lambda\rho}
	+\eta_{\mu\lambda }\eta_{\nu \rho} 
	+\eta_{\mu\rho }\eta_{\nu \lambda }, 
\\&&\hskip-25pt 
	T_{\mu\nu\lambda\rho}
	=\eta_{\mu\nu}q_{\lambda}q_{\rho}
	+\eta_{\mu \rho }q_{\lambda }q_{\nu }
	+\eta_{\nu\lambda }q_{\mu }q_ {\rho }
	+\eta_{\lambda\rho }q_{\mu }q_{\nu },
\\&&\hskip-25pt 
	T'_{\mu\nu\lambda\rho}
	=\eta_{\mu\lambda }q_{\nu }q_{\rho}
	+\eta_{\nu\rho }q_{\mu }q_ {\lambda }.
\end{eqnarray}

We substitute (\ref{Iresult})--(\ref{Jresult}) 
	into (\ref{G^1=}) and (\ref{G^2=}),
	and substitute them into (\ref{FourierGl}) to get $ {\cal L}^{\rm eff}$,
	and rearrange the terms into a sum of monomials of 
	$h_{\mu\nu}$, $A^m{}_{n\mu}$, $B_{m\mu\nu}$ and their derivatives.
Each term is proportional to ${\cal I}_j M_j^{2k}$ ($k=0,1,2$), 
	which, when regularized via (\ref{Sreg}), behave as
\begin{eqnarray}&& 
	\sum_j C_j {\cal I}_j \rightarrow \frac{\ln\Lambda^2}{(4\pi)^2},\ \ 
\cr&&
	\sum_j C_j {\cal I}_j M_j^2 \rightarrow -\frac{ \Lambda^2}{2(4\pi)^2},\ \ 
\cr&&
	\sum_j C_j {\cal I}_j M_j^4 \rightarrow \frac{ \Lambda^4}{2(4\pi)^2},\ \ 
\end{eqnarray}
for large $\Lambda$ (=the equal mass limit of $M_j$). 
The terms are classified as follows.
\\(i) The terms with $N_A=N_B=0$ are given by \cite{h^superscripts}
\begin{eqnarray}&& 
	\frac{ N_{\rm ex}}{32 (4\pi)^2} \bigg[
	\frac{ \Lambda^4}{2} (4h-2h_{(2)} +h^2)
\cr&&
	+\frac{ \Lambda^2}{3} 
	(h^{\mu\nu,\lambda} h_{\mu\nu,\lambda}
	-2h^{\mu\nu}{}_{,\nu} h_{\mu}{}^{\lambda}{}_{,\lambda}
	+2h^{\mu\nu}{}_{,\nu} h_{,\mu}
	-h^{,\mu} h_{,\mu})
\cr&&
	+\frac{ \ln\Lambda^2}{15} 
	\big \{ h^{\mu\nu,\lambda\rho} h_{\mu\nu,\lambda\rho}
	-2h^{\mu\nu}{}_{,\nu\rho } h_{\mu}{}^{\lambda}{}_{,\lambda\rho }
\cr&&\hskip20pt
	+4(h^{\mu\nu}{}_{,\mu\nu})^2 
	-6h^{\mu\nu}{}_{,\mu\nu} h^{,\lambda}{}_{\lambda} 
	+3(h^{,\mu}{}_{\mu})^2\big\} \bigg] 
  \label{hhh}
\end{eqnarray}
up to total derivatives.
Because the full expression should have the symmetry, 
	they should be the lower order expression of 
	$\sqrt{-g}$ times the invariant forms in table \ref{tab1}.
The terms in (\ref{hhh}) are to be compared with 
	the lower contributions for $\sqrt{-g}$ in (\ref{sqrt-g-inh}) and
\begin{eqnarray}&& 
	\sqrt{-g}R=-\frac{1}{4} 
	 (h^{\mu\nu,\lambda} h_{\mu\nu,\lambda}
	-2h^{\mu\nu}{}_{,\nu} h_{\mu}{}^{\lambda}{}_{,\lambda}
\cr&&\hskip80pt
	+2h^{\mu\nu}{}_{,\nu} h_{,\mu}
	-h^{,\mu} h_{,\mu}),
\\&&
	\sqrt{-g}R^2=
	(h^{\mu\nu}{}_{,\mu\nu})^2 
	-2h^{\mu\nu}{}_{,\mu\nu} h^{,\lambda}{}_{\lambda} 
	+(h^{,\mu}{}_{\mu})^2,
\\&&
	\sqrt{-g}R_{\mu\nu}R^{\mu\nu}=-\frac{1}{4} 
	\big [ h^{\mu\nu,\lambda\rho} h_{\mu\nu,\lambda\rho}
	-2h^{\mu\nu}{}_{,\nu\rho } h_{\mu}{}^{\lambda}{}_{,\lambda\rho }
\cr&&\hskip20pt
	+2(h^{\mu\nu}{}_{,\mu\nu})^2 
	-2h^{\mu\nu}{}_{,\mu\nu} h^{,\lambda}{}_{\lambda} 
	+ (h^{,\mu}{}_{\mu})^2\big],
\end{eqnarray}
where total derivatives are neglected.
\\(ii) The lowest contributions to ${\cal L}^{\rm reg}$ 
	with $N_B\not=0$ and $N_\partial=0$ are 
\begin{eqnarray}&&\hskip-10mm
	\frac{1}{4(4\pi)^2} \left(
	{\Lambda^2}B^{(2)} 
	+{\ln\Lambda^2}B^{(4)}	\right),
\end{eqnarray}
which are taken as the lowest parts of the forms 
	$\sqrt{-g} B^{(2)}$ and $\sqrt{-g} B^{(4)}$.
\\(iii) The lowest contribution with $N_B\not=0$ and $N_\partial=2$ is
\begin{eqnarray}&&\hskip-10mm
	\frac{\ln\Lambda^2}{12(4\pi)^2} 
	( h^{\mu\nu}{}_{,\mu\nu}+ h^{,\mu}{}_{\mu}) B^{(2)}
\end{eqnarray}
which is the lowest part of the form 
	$\sqrt{-g}R B^{(2)}$.
\\(iv) The lowest contribution with $N_A\not=0$ is 
\begin{eqnarray}&&\hskip -10mm
	\frac{1}{24(4\pi)^2} 
	(A_{mn\mu,\nu}-A_{mn\nu,\mu})(A^{mn\mu,\nu}-A^{mn\nu,\mu}),
\end{eqnarray}
which is the lowest part of the form 
	$ \sqrt{-g}A_{mn\mu\nu} A^{mn\mu\nu}$ with $N_A=2$.
Note that it suffices to determine the coefficient of the form in ${\cal L}^{\rm reg}$.

Collecting the results of (i)--(iv), 
	we finally obtain the expression for 
	the divergent part ${\cal L}^{\rm div}$ of ${\cal L}^{\rm reg}$:
\begin{eqnarray}&&\hskip-10pt 
	{\cal L}^{\rm div}=\sqrt{-g}/(4\pi)^2
\cr&&
	\times\bigg[N_{\rm ex}\left\{
	\frac{\Lambda^4}{8}-\frac{\Lambda^2}{24}R
	+\frac{\ln\Lambda^2}{240}(R^2 + 2R_{\mu\nu} R^{\mu\nu})
	\right\}
\cr&&\hskip10mm 
	+\frac{\Lambda^2}{4}B^{(2)} 
	+\frac{\ln\Lambda^2}{4}B^{(4)}
	-\frac{\ln\Lambda^2}{12}R B^{(2)}
\cr&&\hskip15mm
	-\frac{\ln\Lambda^2}{24}A_{mn\mu\nu} A^{mn\mu\nu}
	\bigg],
  \label{Sdiv}
\end{eqnarray}
where	$B^{(2)}$, $B^{(4)}$ and $ A_{mn\mu\nu}$ are defined in 
	(\ref{BB}), (\ref{BBBB}) and (\ref{Amnmunu}), respectively,
	and $ N_{\rm ex}$ is the number of the extra dimensions.
The divergences cannot be renormalized 
	because the original action does not have these terms.
They give rise to genuine quantum induced effects.

\section{Cosmological Constant \label{cosmological}}

Thus, we have derived the quantum effects of the brane fluctuations.
Among them, the $\Lambda^4 $ term in (\ref{Sdiv}) gives 
	huge a contribution to the cosmological term.
To this term, the starting Lagrangian ${\cal L}'_{\rm br}$ 
	in (\ref{NG'})	by itself also has a contribution. 
From phenomenological points of view, it should be very tiny.
Therefore, the large contributions should cancel out each other
	to give the tiny cosmological term.
The condition for the cancellation is 
\begin{eqnarray}
	\lambda =- N_{\rm ex}\Lambda^4/128\pi.
  \label{cosmcan}
\end{eqnarray}
This is,  however, an extremely unnatural fine tuning.
It is a serious problem common to the quantum theories including gravity in general.
The present formulation has no solution to this longstanding problem.

Furthermore, it may give rise to another contribution 
	which may mimic the cosmological term 
	in the effective equation of motion for $g_{\mu\nu}$.
The energy momentum tensor in the equation has the term
\begin{eqnarray}
	\lambda Y^I_{,\mu} Y^J_{,\nu} G_{IJ}/2,
\end{eqnarray}
which may look like the cosmological term
	if the embedding is almost flat.
In such cases,
we can adjust the cosmological term to the phenomenological tiny value
	by, for example, adopting the conformally flat embedding
\begin{eqnarray}
	Y^\mu 
	=[1+ N_{\rm ex}\Lambda^4/128\pi\lambda]^{1/2} x^\mu,
	\ \ \  Y^m=0
  \label{confemb}
\end{eqnarray}
instead of the condition (\ref{cosmcan}).
This is also an extremely unnatural fine tuning.
Thus, the present model is not satisfactory 
	in natural understanding of the cosmological constant.
It is,  however, not a problem for the model alone,
	but a serious puzzle for general quantum theoretical models 
	with gravity.
It is an open problem, 
	and we wish that it will be solved in the future.
The problem will be partly addressed in our forthcoming paper.
Here, we phenomenologically adjust the tiny cosmological term
	via fine tuning of (\ref{cosmcan}) or (\ref{confemb}).

\section{Induced gravity \label{induced gravity}}

If the cosmological term is suppressed, 
	the main contribution in the quantum effects (\ref{Sdiv})
	comes from the $R$ term.
It is nothing but the Einstein-Hilbert action,
	which supply the kinetic term for the auxiliary field $g_{\mu\nu}$.
The sign of the term is right one to give ordinary attractive gravity 
	in accordance with the observation,
	and its magnitude indicates 
	that the cutoff $\Lambda$ is order of the Planck scale.
The term with $(R^2 + 2R_{\mu\nu} R^{\mu\nu})$
	gives small corrections of $O(\log\Lambda^2/\Lambda^2)$
	as far as the brane curvature is small.
The terms with $B^{(2)}$, $RB^{(2)}$and $B^{(4)}$ 
	are the mass and interaction terms
	of the field $B_{m\mu\nu}$. 
Note that no kinetic term for $B_{m\mu\nu}$ appears.
This is because the $B_{m\mu\nu}$ interacts with $\phi^m$
	only in the combination $ B_{m\mu\nu} B^{n\mu\nu}$,
	but not in single.
The term with $A_{mn\mu\nu}$ squared gives 
	the kinetic and the interaction terms of the field $A_{mn\mu}$ 
	as the gauge field. 
The fields $A_{mn\mu}$ and $B_{m\mu\nu}$ appear as fields on the brane.
We should, however, be careful because
	they are not independent  
	and are defined by (\ref{Amnmu}) and (\ref{Bmmunu})
	in terms of $Y^I$ and $n^m$.

The quantum induced terms in (\ref{Sdiv}) modify the equations of motion.
The equation (\ref{NG'EM}) for $Y^I$ is modified 
	through the $B^{(2)}$, $RB^{(2)}$, $B^{(4)}$ 
	and $(A_{mn\mu\nu})^2$ terms in (\ref{Sdiv}).
The classical solution $Y^I$ is deformed according to it.
The correction terms are suppressed by at least a factor of $O(\Lambda^{-2})$
	for small curvatures.
{The equation} (\ref{gmunu})
	{for $g_{\mu\nu}$, the induced-metric formula, is converted 
	into the Einstein equation} with 
	the $O(R^2)$ correction terms 
	and the energy momentum tensor for the fields 
	$\phi^m$, $A_{mn\mu}$, $B_{m\mu\nu}$, and $Y^I$. 
The equation (\ref{gmunu}) holds as operator relation.
In classical realizations, however, 
	it suffers from a large quantum corrections.
Then, we no longer have the induced-metric formula.

\section{Discussions \label{discussions}}

The metric $g_{\mu\nu}$ emerges in the channel of intermediate quantum states
	composed of $\phi^m$'s,
	despite its absence in the original setup of the system (\ref{NG}).
Hence, it is interpreted as a composite of 
	the brane fluctuation fields $\phi^m$. 
Then the natural question is what is the further quantum effects
	of the composite metric.
Within the semi-classical treatments, 
	it suffices to calculate only the one loop diagrams.
The system does not include multi-loop diagrams.
This is the virtue of the linear definition (\ref{fluctuation}) 
	of the brane fluctuation.
Beyond the semi-classical approximation, however,
	we should take into account the higher order diagrams
	with the internal lines of composite metric fields.

The quantum induction mechanism of composite fields
	is common phenomena to various composite field theories \cite{comp}.
A class of non-renormalizable theories with this mechanism becomes 
	equivalent to some remormalizable models (with finite momentum cutoff)
	under the ``compositeness condition" 
	that the wave-function renormalization constant vanishes
	\cite{compcond}, \cite{Shizuya}.
This renders us clues to formulate unambiguously
	the non-renormalizable theories at higher orders \cite{compcond2}.
For example, the Nambu-Jona-Lasinio model is equivalent to 
	the Yukawa model with the vanishing renormalization constants
	of the scalar and the pseudoscalar fields,
	and the latter renders a unambiguous higher-order descriptions of the former.

In the present case, however, 
	the induced composite field theory is the modified Einstein gravity,
	and is not renormalizable. 
We have no definite way to calculate the quantum effects 
	due to the metric itself at higher orders.
It shares the problems with the general quantum gravity theories.
So we cannot apply all the achievements of the composite field theories
	with the compositeness condition.
They are, however, very suggestive
	in considering properties of the quantum fluctuations.
In composite theories, it is plausible that
	the quantum effects due to the composite 
	would require different treatments.
For example, if the cutoff for the composites is much smaller 
	than that for the constituents, 
	the effects can be suppressed \cite{compcutoff}.
Or, the 1/$N$ expansion would be useful, as is in the various composite field
	theories \cite{Shizuya}, \cite{compcond2}, \cite{Tomboulis}.
We need further ideas and investigations for the more complete treatments.

We can see in (\ref{Sdiv}) that the quantum effects give rise to 
	the terms including the extrinsic curvature $B_{m\mu\nu}$
	and the normal connection $A_{mn\mu}$, 
	in addition to the induced-gravity terms
	\cite{GibbonsSuggest}.
The induction of these terms 
	is characteristic of the brane induced gravity theory,
	distinguished from the ordinary (non-brane) induced gravity.
The fact was recognized in \cite{Akama87} and \cite{Akama88}
	in the general braneworld scheme, 
	and they were actually calculated in \cite{Akama88b}, \cite{AH}
	for the domain-wall type braneworld. 
The forms of induced terms depend on the brane dynamics.
The simplest case of the Nambu-Goto action was considered in \cite{Akama78}
	within the four dimensional field theory.
In the model, however, only the gravity is induced, but no other terms. 
This is because the spacetime spanned by the scalar fields is not the real one, 
	and hence it assumes no symmetry of the whole spacetime 
	involving the brane.
Therefore, we cannot define the normal to the brane.  
On the contrary, the present model (\ref{NG}) possesses 
	general-coordinate invariance of the bulk, as well as that of the brane.
Therefore, the fluctuations along the brane is meaningless, and 
	the only physical ones are those transverse to the brane,
	as are defined by (\ref{nY})--(\ref{YY+nn=G}).
This is the origin how it includes 
	the $A_{mn\mu} $ and $B_{m\mu\nu}$ dependent terms.
They should be determined according to the brane dynamics,
	as is done here.
It would be an interesting and urgent subject
	to derive the induced terms in various brane dynamics,
	and seek for the models suited for applications.

\begin{acknowledgments}

We would like to thank 
Professor G.~R.~Dvali,
Professor G.~Gabadadze,
Professor M.~E.~Shaposhnikov, 
Professor I.~Antoniadis, 
Professor M.~Giovannini,
Professor S.~Randjbar-Daemi, 
Professor R.~Gregory, 
Professor P.~Kanti, 
Professor G.~Gibbons, 
Professor K.~Hashimoto, 
Professor E.~J.~Copeland, 
Professor D.~L.~Wiltshire, 
Professor I.~P.~Neupane,
Professor R.~R.~Volkas, 
Professor A.~Kobakhidze, 
Professor C.~Wetterich, 
Professor M.~Shifman, 
Professor A.~Vainshtein, 
Professor D.~Wands, 
Professor M.~Visser, 
Professor T.~Inami, 
Professor I.~Oda and 
Professor H.~Mukaida 
for invaluable discussions and their kind hospitality extended to us
during our stay in their places.

This work was supported by Grant-in-Aid for Scientific Research,
No.\ 13640297, 17500601, and 22500819
from Japanese Ministry of Education, Culture, Sports, Science and Technology.

\end{acknowledgments}



\begin{thebibliography}{99}
\bibitem{GibbonsSuggest}
We would like to thank Professor G. Gibbons 
	for having suggested us the existence of these extrinsic effects
	in the present model.
\bibitem{h^superscripts}
Note that $h_{\mu\nu}$ and its derivatives are not brane tensors, 
	and their superscripts are not defined by multiplying $g^{\mu\nu}$.
For convenience of calculations, 
	we define here that their subscripts are raised by $\eta^{\mu\nu}$.
\end{thebibliography}
\end{document}